\documentclass[12pt]{article}
\usepackage{amsmath}
\usepackage{amssymb}
\usepackage{a4wide}
\numberwithin{equation}{section}

\def\defeq{\,\raisebox{0.4pt}{:}\hspace{-1.2mm}=}
\def\pare#1{\left( #1\right)}
\def\bpare#1{\left\{ #1\right\}}

\def\rep#1{\mathbf{#1}}
\def\bZ{\mathbb{Z}}
\def\tr{\mathop{\mathrm{tr}}\nolimits}
\def\trad{\tr_{\mathrm{ad}}}

\def\gaugeheading#1{\bigskip
\noindent$\bullet$\ \, #1\ :\ \ }
\def\lie#1{\mathsf{#1}}

\begin{document}

\begin{titlepage}
\begin{flushright}
UT-05-17\\
NSF-KITP-05-100\\
hep-th/0512019
\end{flushright}

\vskip 1cm

\begin{center}
\textbf{\LARGE More anomaly-free models of \\[4mm]
six-dimensional gauged supergravity}\\

\vskip1cm

{\Large Ryo Suzuki$^1$ and Yuji Tachikawa$^{1,2}$}\\

\vskip1cm 

\textit {$^1$\ {\large Department of Physics, Faculty of Science,}}\\
\textit {\large{University of Tokyo, Tokyo 113-0033,  JAPAN}}\\

\vskip.5cm

\textit {$^2$\ \large{Kavli Institute for Theoretical Physics,}}\\
\textit {\large{University of California, Santa Barbara, California 93106 , USA}}\\

\vskip .7cm

e-mails: \texttt{ryo,yujitach@hep-th.phys.s.u-tokyo.ac.jp}\\

\vskip1.7cm

{\large\textbf{abstract}}
\end{center}

\vskip.5cm

We construct a huge number of anomaly-free models of 
six-dimensional ${\cal N} = (1,0)$ gauged supergravity.
The gauge groups are products of $U(1)$ and $SU(2)$, and
every hyperino is charged under some of the gauge groups.
It is also found that the potential may have flat directions when the $R$-symmetry is diagonally gauged together with another gauge group.
In an appendix, 
we determine the contribution to the global $SU(2)$ anomaly
from symplectic Majorana Weyl fermions in six dimensions.
\vbox{}\vspace{1\fill}

\end{titlepage}

\newpage

\section{Introduction}

Six-dimensional ${\cal N}=(1,0)$ supergravity \cite{NS86,NS97} has many interesting features.
The ungauged version has been useful in uncovering
the interesting dynamics of string theory in six dimensions.
The gauged one is particularly interesting,
because it does not allow the flat six-dimensional Minkowski spacetime as a solution. 
Their solutions typically describe
spacetimes which are spontaneously compactified to lower dimensions \cite{SS84,GLPS03}.
They can also be used to build various higher-dimensional models of
particle phenomenology and cosmology. See, e.g., \cite{ABCFPQTZ03,BR04}.

Any higher dimensional theory of gravity should be considered
as a low energy approximation of some unknown quantized theory,
and there are several consistency conditions that any low energy approximation should satisfy. Anomaly freedom is one of the most important criteria.
The search for anomaly-free models in six dimensions is more difficult and
at the same time richer than in ten dimensions.
It is because in six dimensions we can include hypermultiplets,
which contribute to perturbative anomalies \cite{AW83}.

The $d=6,\, {\cal N}=(1,0)$ ungauged supergravity
can be obtained from heterotic strings on $K3$, and
many anomaly-free models are known \cite{Ketov88,Ketov90,GSW85,Schwarz95,SW96}, with the help of
the Green-Schwarz mechanism \cite{GS84} in six dimensions.
For $d=6,\, {\cal N}=(1,0)$ gauged supergravity, however,
only a handful of consistent models have been found so far.
Furthermore, if we impose the constraint that all hyperini should be charged
under some of the gauge groups, the number of consistent models is very small \cite{RSSS85,AKR05,AK05}.

In $d=10,\, {\cal N}=1$ supergravity the anomaly cancels only for a few models,
namely $SO(32), E_8 \times E_8, E_8 \times U(1)^{248}$
and $U(1)^{496}$.
Moreover,
the discovery of anomaly freedom of $E_8 \times E_8$
inspired the construction of heterotic string theories. 
It is thus quite interesting to study 
how many anomaly-free models there are in 
$d=6,\, {\cal N}=(1,0)$ gauged supergravity,
and it might suggest the existence of
some totally novel quantum completion of those theories
within superstring theory or outside of it.
No consistent way to derive it from the compactification of string or $M$ theory is not known yet,
although some progress is being made \cite{CGP03,BJS05}.
This is also interesting from the point of view of its phenomenological or cosmological applications.

In this paper,
we investigate the models whose gauge groups are
products of $U(1)$ and $SU(2)$.
This choice makes the condition for anomaly cancellation relatively simple.
It will be shown that there are enormously many models 
which are free of  both perturbative and global anomaly.

\bigskip

The paper is organized as follows.  First we recall
basic knowledge on $d=6,\, {\cal N}=(1,0)$ gauged supergravity in section 2.
In section 3, we describe the general form of both perturbative and
global anomaly-free conditions, and 
we carry out the search and give our results in section 4.
Section 5 is the summary and discussion.
In appendix A we collect our notations concerning the group representations.
Appendix B discusses the global gauge anomaly
from the symplectic Majorana Weyl fermion charged under $SU(2)$ gauge group.

\section{Gauged ${\cal N}=(1,0)$ supergravity in six dimensions}

\subsection{The spectrum}\label{sec2.1}

${\cal N}=(1,0)$ supergravity in six dimensions contains the following multiplets:
\begin{align*}
\text{supergravity multiplet}, \hspace{15mm}
&( e_\mu^{\;\; m} , B^-_{\mu\nu} \,, \psi_\mu^{\;\;A-} )\,; \\
\text{tensor multiplet}, \hspace{15mm}
&( B^+_{\mu\nu} \,, \chi^{A+} , \varphi )\,; \\
\text{vector multiplet}, \hspace{15mm}
&( A_\mu \,, \lambda^{A-} )\,; \\
\text{hypermultiplet}, \hspace{15mm}
&( 4 \, \phi , \, \psi^+ )\,; 
\end{align*}
where $\mu, \nu = 0, \ldots, 5$ label spacetime, 
$m = 0, \ldots, 5$ labels tangent space, 
$A = 1, 2$ labels the fundamental representation of $Sp(1)_R$\,, 
and $\pm$ denotes the chirality of Weyl spinors 
or the self-duality of the field strength of antisymmetric two-forms.

Weyl spinors of $SO(1,5)$ and the fundamental representation of $Sp(1)_R$ are both pseudoreal. 
By combining two antilinear involutions, 
we can impose a reality condition to get symplectic Majorana Weyl spinors. 
Gravitini, tensorini and gaugini are symplectic Majorana Weyl under $Sp(1)_R$.
Hyperini are inert under $Sp(1)_R$ and are Weyl spinors in general. 
If some of the hypermultiplets form a pseudoreal representation under the gauge groups, 
then we can impose the symplectic reality condition on them. 
Such a hypermultiplet is called a half-hypermultiplet.

Hereafter we assume the number of tensor multiplet $n_T$ is one. 
This is because only in this case Lorentz- and gauge-covariant Lagrangian exist at the classical level.

\subsection{Gauging hyperscalar manifold}

The $d=6, {\cal N}=(1,0)$ rigid supersymmetry requires the scalar fields in the
hypermultiplets to parametrize a hyperk\"{a}hler manifold. 
If we couple hypermultiplets to gravity, then it must be
a quaternionic manifold with negative curvature. 
For simplicity we assume the target space of the hyperscalars to be the manifold
\begin{equation}
{\cal M}_H \defeq \frac{Sp(1,n_H)}{Sp(1)_R \times Sp(n_H)} \,,
\label{qhk}
\end{equation}
where $n_H$ is the number of hypermultiplets.

We introduce vector multiplets which  gauge 
some part of the isometry group $Sp(1)_R \times Sp(n_H)$ of ${\cal M}_H$.
Arbitrary subgroups of $Sp(1)_R \times Sp(n_H)$ can be gauged.
Let us write the gauge groups as $G_R\times G_H\subset Sp(1)_R\times Sp(n_H)$, 
where $G_R\subset Sp(1)_R\times Sp(n_H)$ gauges some part of the
$R$-symmetry and $G_H\subset Sp(n_H)$ acts only on the hypermultiplets. 
The closure of Lie algebra requires that $G_R$ be one of the following possibilities:
$U(1)_R$, $Sp(1)_R$, $U(1)_{R+}$ and $Sp(1)_{R+}$\,, where
in the latter two cases we take $U(1)$ or $Sp(1)$ subgroup of $Sp(n_H)$
and gauge the diagonal combination of them with $U(1)_R$ or $Sp(1)_R$, respectively.
We call these latter two choices 
as `diagonal gaugings.'
When $G_H$ is made out of several factors $G_{H1} \times G_{H2} \times \cdots$,
we use the label $z = 1, 2, \ldots$ to distinguish different factors.

Gauging hyperscalar manifold brings an additional potential term to the Lagrangian, 
which is required by supersymmetry.
Here we will write the general form of the potential, citing the results of \cite{NS86,NS97}.
We denote hyperscalars parameterizing the manifold ${\cal M}_H$ by $\phi^\alpha,\ \alpha \, = \, 1, \ldots, \, 4 \, n_H$\,.
Since ${\cal M}_H$ is a symmetric space, 
its tangent space is spanned by the coset of the Lie algebras. 
Let $L (\phi)$ be a representative of the coset $Sp(1,n_H) / [ Sp(1)_R \times Sp(n_H) ]$
so that $L (\phi) = {\bf 1}_{Sp(1,n_H)} + {\cal O} (\phi)$,
and define $C$-functions as\footnote{The $C$-function is known under various names: 
$P$ function, Killing prepotential, triholomorphic moment map, etc.}
\begin{equation}
C^{\, AB}_{\ \ \ \ CD} \defeq \big( \, L^{-1} T^{AB} L \, \big)_{CD} \,, 
\quad  C^{\; ab}_{z \;\;\;\; CD} \defeq \big( \, L^{-1} T_z^{ab} L \, \big)_{CD} \,,
\label{c-func1}
\end{equation}
where $T^{AB}$ and $T_z^{ab}$ are the generators of $G_R$ and $G_{Hz}$\,, respectively. 
Then, the potential is given by
\begin{equation}
V (\phi) = \frac14 \, e^{-\varphi} \bigg( \, g_R^2 \,  
C^{AB}{}_{CD} \, C^{\, AB \, CD} + \sum_z \, g_{Hz}^2 \,
C_z^{\; ab}{}_{CD} \, C_z^{\; ab \, CD} \, \bigg) 
\label{potential1}
\end{equation}
where $g_R \,, g_{Hz}$ are the coupling constants\footnote{We normalize the gauge kinetic term as $- \frac{e^{\pm \varphi}}{4 \, g_k^2} \tr_k F_{\mu\nu} F^{\mu\nu}$, with $k=R, H_z$. 
}  of $G_R \,, G_{Hz}$.

The potential \eqref{potential1} is nonnegative, because it is the sum of the squares of $C$-functions.
One important feature is that $C^{AB}{}_{CD}=T^{AB}{}_{DC}+{\cal O}(\phi)$ and $C_z^{\; ab}{}_{CD}={\cal O}(\phi)$,
hence the potential is positive at $\phi^\alpha=0$ if we gauge $R$-symmetry.
It provides a positive cosmological constant in a six-dimensional sense. 
For a non-diagonal gauging, 
explicit calculations of \cite{NS86,AKR05,RS04} show that 
$\phi^\alpha=0$ is the minimum of the potential and there is no possibility of Higgsing.
If we take a diagonal gauging, however,
the potential may have flat or tachyonic directions.
In this case, one can spontaneously break $R$ symmetry and 
possibly it leads to ungauged supergravity theories. 
The physics of models with diagonal gauging is relatively unexplored,
and we hope to revisit this problem in the future.

Another important feature is that in the examples discussed so far, 
quadratic terms of $V (\phi)$ are only from the $R$ coupling. 
Thus, the mass of hyperscalars is determined by their $R$ charges.

The way how hyperini acquire four-dimensional mass depends on the details of compactification.
Consider for example, a spacetime $\mathbb{R}^{1,3} \times S^2$
with monopoles in the internal $S^2$. 
If we embed the field strength of the monopole in $G_R = U(1)_R$, 
 $d=4$, ${\cal N}=1$ supersymmetry can remain unbroken \cite{SS84}. 
Other choices of monopole charges generically break all the supersymmetry and many of them induce instability\footnote{See \cite{RSS82} for  the models with monopoles sitting in the $U(1)$ factor, and \cite{RSS83} for those with monopoles sitting in the nonabelian factor.}.

\section{General anomaly-free conditions}

Any six-dimensional gauge theory must satisfy two constraints 
concerning its gauge groups and representations. They are the freedom from
the local and the global gravitational, gauge and mixed anomaly.
The local or the global anomaly measures
the change in the fermion determinant, induced by a
gauge transformation which can or cannot be
continuously deformed to identity. 
One must choose the gauge groups and the representations carefully
so that both kinds of anomaly will cancel.

\subsection{Local anomaly}

It is well-known that the Green-Schwarz mechanism 
can cancel the local gravitational, gauge and mixed anomaly if anomaly polynomial factorizes.\footnote{Note that in six dimensions the consistency of the Green-Schwarz mechanism is rather subtle, 
because we need to modify the lowest-derivative terms in the Lagrangian 
in order to introduce the Green-Schwarz counterterm.
More details can be found in references \cite{NS97,FMS96}, 
including the generalization to  $n_T > 1$.}

Anomaly polynomial can be explicitly calculated
by summing up the contributions from fermions and (anti-)selfdual tensors. 
Fermions of positive chirality or antisymmetric tensor 
with self-dual field strength contribute to it positively, 
while fermions of negative chirality or antisymmetric tensor 
with anti-self-dual field strength do negatively. 
Hence in our case, the total anomaly polynomial $P_{\rm total}$ is of the form
\begin{equation}
P_{\rm total} = - \pare{I_{3/2} + I_A} + \sum_{\rm tensor} \pare{I_A + I_{1/2}} - 
	\sum_{\rm vector} I_{1/2} + \sum_{\rm hyper} I_{1/2} \,,
\label{ptotal}
\end{equation}
where each term comes from the supergravity multiplet, 
tensor multiplets, vector multiplets and hypermultiplets, respectively.

In six dimensions, the anomaly polynomials for spin 3/2 fermions and spin 1/2 fermions in the representation $r$ are known to be \cite{AW83,AG84}
\begin{align}
I_{3/2} &= \left(\frac{245}{360} \,\tr R^4 - \frac{43}{288} \,\pare{\tr R^2}^2\right)\tr_r 1 + 
\frac{19}{6} \,\tr R^2 \,\tr_r F^2 + \frac{10}{3} \,\tr_r F^4 ,
\label{I_3/2}   \\
I_{1/2} &= \left(\frac{1}{360} \,\tr R^4 + \frac{1}{288} \,\pare{\tr R^2}^2\right)\tr_r 1
 - \frac{1}{6} \,\tr R^2 \,\tr_r F^2 + \frac{2}{3} \, \tr_r F^4 ,
\label{I_1/2}
\end{align}
and that for a real self-dual antisymmetric tensor to be
\begin{equation}
I_{A} = \frac{28}{360} \,\tr R^4 - \frac{8}{288} \,\pare{\tr R^2}^2 .
\label{I_A}
\end{equation}

We will study what conditions are necessary for \eqref{ptotal} to factorize into the product of four-forms.

First of all, the coefficients of $\tr R^4$ and $\tr_r F^4$ must vanish. Using \eqref{I_3/2} and \eqref{I_1/2}, the $\tr R^4$ condition gives
\begin{equation}
n_H = 273 - 29 \, n_T + n_V \,.
\end{equation}
To satisfy the $\tr_r F^4$ condition, we restrict our analysis 
to particular representations 
for which $\tr F^4$ is a multiple of $\pare{\tr F^2}^2$.
We call such representations `exceptional-type.' 
All finite dimensional irreducible representations
of $\lie{A}_1$, $\lie{A}_2$, $\lie{E}_6$, $\lie{E}_7$, $\lie{E}_8$, $\lie{F}_4$ and 
$\lie{G}_2$ are of exceptional type, which we call Lie algebras of exceptional type. 
Further studies on exceptional-type representations can be found in
\cite{Okubo78,Okubo81}.
Casimir invariants of exceptional-type Lie algebras are summarized in appendix \ref{appA}.

When $\tr R^4$ and $\tr_r F^4$ vanish, we can rewrite the total anomaly polynomial as
\begin{equation}
P_{\rm total} = \sum_{jk} \beta_{jk} \, K^j K^k \,.
\label{pbeta}
\end{equation}
where $K^k$ is
\begin{equation}
\vec K \defeq (\tr  R^2, \tr _f F_{G_R}^2, \tr _{f} F_{G_1}^2, \tr _{f} F_{G_2}^2, \ldots, \tr _{f} F_{G_n}^2)
\end{equation}
where $G_i$ is shorthand notation for $G_{Hi}$
and $f$ is the smallest nontrivial irreducible representation of $G_i$.

It is convenient to regard $\beta_{ij}$ as  a $(n+1) \times (n+1)$ matrix. 
We call $\beta_{ij}$ anomaly matrix of the model.
The condition for factorization of $P_{\rm total}$ is
equivalent to $\beta_{ij} = \pare{\alpha_i \gamma_j + \alpha_j \gamma_i}/2$ 
for some $\alpha_i, \gamma_j$\,. 
It is clear that the columns of $\beta$ are linear combination of
$\vec \alpha$ and $\vec \gamma$, so we must have
\begin{equation}
{\rm rank}\, \beta \le 2.  \label{pmatfirst}
\end{equation}
Besides, the two vectors $\vec \alpha$ and $\vec \gamma$ must be real, because they enter into Lagrangian through the Green-Schwarz counterterm.
Elementary calculation shows that $\beta$ has two real non-zero eigenvectors (and thus real $\vec \alpha$ and $\vec \gamma$) if and only if
\begin{equation}
\lambda^+ \lambda^- \le 0. \label{pmatcond}
\end{equation}
We call \eqref{pmatfirst} and \eqref{pmatcond} the first and the second factorization condition, respectively.

\bigskip

As a preparation for the actual search for anomaly-free models in the next section, we describe the anomaly matrix more explicitly in each case of $G_R = U(1)_{R(+)}$ and $Sp(1)_{R(+)}$\,. Throughout the paper, 
we write the representation of hyperini as $\rho^H$ and use $\tr_H$ as the abbreviation for the trace over $\rho^H$.

For $G_R = U(1)_R$\,, the gravitini, tensorini and gaugini all have
charge one under $U(1)_R$. The gaugini are the adjoints
of the gauge groups.
The anomaly polynomial is given by\footnote{We normalize the total anomaly polynomial so that $\alpha_1 = \gamma_1 = 1$}
\begin{align}
P &= \pare{\tr R^2}^2  + \frac{\tr R^2}{6} \pare{
		\pare{- 20 + n_V} \, F_{U(1)R}^2 + \sum_{i=1}^n \trad  F_{G_i}^2    - \tr_{H} F^2}   \notag \\
&+ \frac23 \bpare{
	- \pare{4 + n_V} \, F_{U(1)R}^4  
	- \sum_{i=1}^n \trad  F_{G_i}^4 
	- 6 \, F_{U(1)R}^2 \sum_{i=1}^n \trad  F_{G_i}^2 + \tr_{H} F^4 
} \, .
\label{ap:3}
\end{align}

For $G_R = U(1)_{R+}$\,,
the anomaly polynomial is almost the same as \eqref{ap:3},
except that hyperini are charged under $U(1)_{R+}$ in this case.

For $G_R = Sp(1)_R$\,, recall that the symplectic Majorana Weyl condition is 
imposed on gravitini, tensorini and gaugini.
Another important point to notice is that the gaugini of
the $Sp(1)_R$ symmetry transforms in the $\mathbf{2}\otimes\mathbf{3}$ representation.
The anomaly polynomial is given by
\begin{align}
P &= \pare{\tr R^2}^2 + \frac{\tr R^2}{6} \pare{ \pare{\frac{- 12 + n_V}{2}} \, \tr_2 F_{Sp(1)}^2
 + \sum_{i=1}^n \trad  F_{G_i}^2
  - \tr_{H} F^2 } \notag \\
&+ \frac23 \bpare{ 
	- \pare{\frac{84 + n_V}{4}} \pare{\tr_2 F_{Sp(1)R}^2}^2  
	- \sum_{i=1}^n \trad  F_{G_i}^4 
        - 3 \, \tr_2 F_{Sp(1)R}^2 \sum_{i=1}^n \trad  F_{G_i}^2 + \tr_{H} F^4
   } \, .
\label{ap:4}
\end{align}

For $G_R = Sp(1)_{R+}$\,, 
we need to take it into account that
hyperini are charged under $Sp(1)_{R+}$\,.

\subsection{Global anomaly}
Once one finds a perturbatively anomaly-free model,
one needs to check whether the global anomaly vanishes.
Global gauge anomaly in six dimensions
may appear if the gauge group $G$ has the nonvanishing sixth homotopy group,
$\pi_6 (G) \neq 0$ \cite{Witten82}.
There are three simple Lie groups with $\pi_6 (G) \neq 0$, namely
$\pi_6(SU(2))=\bZ_{12}$\,, 
$\pi_6(SU(3))=\bZ_{6}$ and 
$\pi_6(\lie{G}_2)=\bZ_3$\,. 
Abelian gauge groups do not cause global anomaly because $\pi_6(U(1))=0$.

The conditions for the cancellation of global gauge anomaly have been investigated through the works of \cite{EN84,Kiritsis86,ZOT87,Tosa,BV97}
for the case of Weyl spinors.
The conditions with symplectic Majorana Weyl spinors 
in six dimensions seem to be absent in the literature, 
so we will give the derivation in appendix \ref{appB}.
The results are
\begin{alignat}{11}
& 1 - 4 \, C_4 \pare{\rho^H; \lie{G}_2} & &\equiv 0 &\quad 
&\pmod{3} &\qquad &{\rm for} &\ &\lie{G}_2\,, & \label{g_anm1}\\
& 8 - \phantom{4} D_4 \pare{\rho^H; SU(2)} & &\equiv 0 &\quad &\pmod{12} 
&\qquad &{\rm for} &\ &SU(2), & \label{g_anm2}\\
& \phantom{8} - 2 \, C_4 \pare{\rho^H; SU(3)} & &\equiv 0 &\quad &\pmod{6} &\qquad &{\rm for} &\ &SU(3) \,, & \label{g_anm3}\\
& n_V - D_4 \pare{\rho^H; Sp(1)} & &\equiv 0 &\quad &\pmod{12} 
&\qquad &{\rm for} &\ &Sp(1)_{R(+)}\,. & \label{g_anm4}
\end{alignat}
where the quantity $C_4$ is
defined in appendix \ref{appA} and the quantity $D_4$ for $SU(2) \simeq Sp(1)$ is defined in appendix \ref{appB}. 
If there are no half-hypermultiplets, the relation
$D_4\equiv 4\,C_4 \pmod{12}$
holds. 
Then the condition \eqref{g_anm2} reduces to
\begin{equation}
4 - 2 \, C_4 \pare{\rho^H; SU(2)} \equiv 0 \pmod{6}, \label{g_anm5}\\
\end{equation}
which is precisely the condition found in \cite{BV97}.

Assuming the vanishing of global gauge anomaly, 
one can show that there is no global gravitational anomaly in six dimensions if the spacetime is $S^6$,
by slightly generalizing the argument in \cite{BKSS86}. 
Furthermore, it means that 
any six-dimensional theory is free of global anomaly on a coordinate patch,
because any large diffeomorphism or large gauge transformation 
on a small patch can be done likewise on $S^6$. 
There might be other global anomalies coming from the 
nontrivial topology of spacetime, but it is beyond the scope of our present work.

\section{Examples of models}

We performed an extensive computer-aided search
of anomaly-free models whose gauge groups
are of the form $G_{R(+)} \times G_H$ where
$G_{R(+)}$ and $G_H$ are $U(1)$ or $SU(2)$.
And then we discovered enormously many anomaly-free models.
In what follows, we describe the details of our search
and show several examples of the models.

\subsection{Abelian gauge groups}

Let $\bpare{h_i}$ be the basis of Cartan subalgebra $u (1)^{n_H}$ of $sp (n_H)$, then the generator of a $u (1) \subset u (1)^{n_H}$ is written as
\begin{equation}
T \defeq \sum_{i=1}^{n_H} q^i \, h_i \,.
\label{u1choice}
\end{equation}
We assume $q^i$'s to be quantized in integers.

When there are more than one abelian factor within the gauge groups
under which hyperini are charged,
the anomaly polynomial, in general, contains terms of the form
\begin{equation*}
\tr F_{U(1)_1} \tr F_{U(1)_2}^3 \,,\quad
\tr F_{U(1)_1} \tr F_{U(1)_2} \tr F_{U(1)_3}^2 \,,\quad
\tr F_{U(1)_1} \tr F_{U(1)_2} \tr F_{U(1)_3} \tr F_{U(1)_4} \,.
\end{equation*}
The presence of traces of odd powers of $F$ necessitates
the generalization of the procedure outlined in the preceding sections. Therefore, in such situations we assume the presence of 
a symmetry among $U(1)$ charges which forbids the appearances of the trace of odd powers of $F_{U(1)_i}$'s.

Before giving our calculation and results, 
let us explain what kind of solutions we seek.
Firstly, if one finds an anomaly-free model,
one can rescale the unit charge of any $U(1)$ and
obtain another solution. This operation is rather trivial,
so we regard two charge vectors related in this way
as the same solution.

Secondly, in the literature, solutions with so-called `drones'
are considered to be unrealistic and uninteresting,
and thus we search for anomaly-free models without drones. 
By drones we mean hypermultiplets which are not charged under $G_R \times G_H$\,, and $U(1)$ vector multiplets with no charged scalars or fermions.

\gaugeheading{$U(1)_R$}
One needs $n_H=245$ neutral hypermultiplets to cancel $\tr R^4$ terms. Then the anomaly polynomial automatically factorizes into \begin{equation}
\Big( \tr R^2 - 4\, F^2_{U(1)R} \Big)
\Big( \tr R^2 +\frac56 \, F^2_{U(1)R} \Big).
\end{equation}
Thus, there is one anomaly-free model, albeit 
lots of singlet hyperini entering into it.

\gaugeheading{$U(1)_R \times U(1)$}
We have found more than 40 million solutions to \eqref{pmatfirst},
\eqref{pmatcond} without drones. Some of them are listed as follows:
\begin{alignat*}{5}
(n_1,n_2,n_3,n_4)\ =\ \ &(243,0,3,0),&\quad &(173,70,3,0),&\quad
&(138,96,12,0),&\quad &(123,102,21,0), \\
&(112,109,24,1),& &(108,96,42,0), & &(108,54,84,0),& &(123,0,123,0),
\end{alignat*}
where $n_q$ is the number of hypermultiplets with charge $q$.
We set $n_q=0$ for $q>4$.

We also found some infinite series of anomaly-free solutions 
with drone $U(1)$ vector multiplets.
For example, if $n_V = 2 + n_{\rm drone}$ and $n_H=246+n_{\text{drone}}$, the combinations
\begin{eqnarray}
\pare{n_1, n_2, n_3, n_4} &=& \pare{243 , 0, 3+ n_{\rm drone}, 0}, \\
\pare{n_1, n_2, n_3, n_4} &=& \pare{173, 70, 3, n_{\rm drone}}
\end{eqnarray}
solve the factorization conditions for any $n_{\rm drone} \ge 1$\,.
There might be a deeper reason why such infinite series exist.

\gaugeheading{$U(1)_{R+}$}
In this case, $\beta_{jk}$ is a two-by-two matrix and the anomaly polynomial immediately factorizes. 
We also need to check the constraint \eqref{pmatcond}, of which one can easily find an enormous amount of solutions with no singlet hypers.

\gaugeheading{$U(1)_{R+} \times U(1)_H$}
Now hyperini can have charges under two abelian groups, so 
the term $\tr F_{U(1)_{R+}} \tr F_{U(1)_{H}} ^{3}$ may appear. 
Let us denote by $n_{ab}$ the number of hyperini 
whose $U(1)_{R+}$ charge is $\pm a$ and whose $U(1)_H$ charge is $\pm b$. 
We restrict $n_{ab}$ to be even so that a half of them have
charge $(a,b)$ and the other half $(a,-b)$. Then the terms containing $\tr F_{U(1)}\,,\tr F_{U(1)}^3$ are removed.

We have found thousands of anomaly-free choices of $n_{ab}$\,,
some of which are:
\begin{equation*}
\begin{pmatrix}
  {n_{11} } & {n_{12} } & {n_{13} }  \\
  {n_{21} } & {n_{22} } & {n_{23} }  \\
  {n_{31} } & {n_{32} } & {n_{33} }  
\end{pmatrix}
=
\begin{pmatrix}
 0 & 0 & 0 \\
 114 & 12 & 2 \\
 22 & 66 & 30
\end{pmatrix},\ 
\begin{pmatrix}
 2 & 4 & 6 \\
 150 & 4 & 2 \\
 6 & 62 & 10 \\
\end{pmatrix},\ 
\begin{pmatrix}
 0 & 0 & 0 \\
 190 & 14 & 30 \\
 10 & 2 & 0
\end{pmatrix},
\end{equation*}
where other $n_{ab}$ are all zero.

\subsection{Non-abelian gauge groups} 

In addition to \eqref{pmatfirst}, \eqref{pmatcond}, one must also check the vanishing of global gauge anomaly for $Sp(1)_R$ and $SU(2)$ when dealing with non-abelian gauge groups. These conditions altogether are quite lengthy and therefore it becomes far rarer to find the solutions than in abelian cases.
Still, we are able to discover hundreds or thousands of anomaly-free models. To be concrete, we will describe some of them in this subsection.

As explained in section \ref{sec2.1},
we can impose symplectic Majorana Weyl condition
to the fermions which transform in a pseudoreal representation.
A half-hypermultiplet contributes to the anomaly polynomial \eqref{I_1/2} half as much as a hypermultiplet.
Thus, once half-hypermultiplets are taken into account,
$n_H$ should be decomposed as
\begin{equation}
n_H =  \sum_r n_r  \dim r \,,
\label{pdim1}
\end{equation}
where $n_r$ is the number of hypermultiplets in the representation $r$,
and we allow $n_r$  to be half-integers if $r$ is pseudoreal.
The group-theoretical constants defined in appendix \ref{appA} then become
\begin{equation}
C_2(\rho^H ;G)=\sum_r n_r  \, C_2(r ;G),\qquad
C_4(\rho^H ;G)=\sum_r n_r  \, C_4(r ;G).
\end{equation}
where $G$ is a non-abelian simple Lie group.

\gaugeheading{$U(1)_R\times SU(2)$}
Anomaly-free choices of $\rho^H$ are listed as follows:
\begin{alignat*}{5}
(n_2, n_3, n_4, n_5, n_6, n_7, n_8)\ =\ \ 
&(0, 4, 1, 11, 26, 3, 0), &\quad
&(0, 7, 0, 2, 0, 31, 0),\\
&(1, 0, 12, 0, 33, 0, 0), &\quad
&(1, 3, 1, 3, 7, 24, 1),\\
&(2, 1, 29, 25, 0, 0, 0), &\quad
&(3, 0, 0, 0, 11, 0, 22),\\
&(5, 0, 0, 0, 37, 0, 2), &\quad
&(124, 0, 0, 0, 0, 0, 0).
\end{alignat*}

\gaugeheading{$U(1)_{R+}\times SU(2)$}
Let $n_{i,r}$ be the number of hypermultiplets with $U(1)_{R+}$ charge $i$ and in the $SU(2)$ representation $r$. 
Let us list some solutions of anomaly-free conditions:
\begin{equation*}
\begin{pmatrix}
n_{1,2} & n_{1,3} & n_{1,4} \\
n_{2,2} & n_{2,3} & n_{2,4} \\
n_{3,2} & n_{3,3} & n_{3,4} 
\end{pmatrix}
=
\begin{pmatrix}
1&0&12\\
66&0&9\\
9&0&3
\end{pmatrix},
\begin{pmatrix}
2&6&9\\
46&4&5\\
28&2&1
\end{pmatrix},
\begin{pmatrix}
3&1&5\\
28&6&1\\
65&1&2
\end{pmatrix},
\begin{pmatrix}
4&5&8\\
59&1&8\\
15&2&1
\end{pmatrix}
\end{equation*}

\gaugeheading{$Sp(1)_R \,, Sp(1)_R \times U(1)$ and $Sp(1)_R \times SU(2)$}
There are no consistent models, because
the $Sp(1)_R$ part has global gauge anomaly.

\gaugeheading{$Sp(1)_{R+}$}
A few examples of anomaly-free spectrum are
\begin{alignat*}{4}
(n_2,n_3,n_4,n_5)\ =\ \ &(107+1/2, 0, 8, 0),&\quad &(109+1/2, 8, 1, 0), \\
&(117+1/2, 1, 1, 1),& &(119+1/2, 0, 2, 0).
\end{alignat*}

\gaugeheading{$Sp(1)_{R+}  \times U(1)$}
Let us denote by $n_{r,i}$ the number of hypermultiplets
with $U(1)$ charge $i$ and in $Sp(1)_R$ representation $r$. 
Hypermultiplets like
\begin{equation*}
\begin{pmatrix}
n_{2,1} & n_{2,2} & n_{2,3} \\
n_{3,1} & n_{3,2} & n_{3,3} \\
n_{4,1} & n_{4,3} & n_{4,3} 
\end{pmatrix}
=
\begin{pmatrix}
0 & 22 & 56 \\
0 & 0 & 0 \\
0 & 4 & 19
\end{pmatrix},
\begin{pmatrix}
15 & 28 & 11 \\
0 & 0 & 0 \\
5 & 11 & 19
\end{pmatrix},
\begin{pmatrix}
23 & 0 & 9 \\
23 & 8 & 1 \\
0 & 18 & 4
\end{pmatrix},
\begin{pmatrix}
32 & 24 & 0 \\
28 & 0 & 4 \\
6 & 2 & 2
\end{pmatrix}
\end{equation*}
give anomaly-free models.

\gaugeheading{$Sp(1)_{R+}  \times SU(2)$}
Let us denote by $n_{r,s}$ the number of hypermultiplets 
in the representation $(r,s)$ of $Sp(1)_R\times SU(2)$. 
Examples of solutions are:
\begin{equation*}
\begin{pmatrix}
n_{1,1} & n_{1,2} & n_{1,3} \\
n_{2,1} & n_{2,2} & n_{2,3} \\
n_{3,1} & n_{3,2} & n_{3,3} 
\end{pmatrix}
=
\begin{pmatrix}
0 & 0 & 0 \\
2 & 6 & 5 \\
26 & 16 & 2
\end{pmatrix},
\begin{pmatrix}
0 & 35 & 0 \\
49 & 7 & 0 \\
9 & 3 & 1
\end{pmatrix},
\begin{pmatrix}
0 & 56 & 1 \\
3 & 0 & 2 \\
12 & 12 & 1
\end{pmatrix},
\begin{pmatrix}
0 & 92 & 5 \\
9 & 6 & 0 \\
0 & 0 & 1
\end{pmatrix},
\end{equation*}
We set $n_{1,1}=0$ to exclude singlet hyperini.

\bigskip

Before closing this section, we would like to mention
an extra anomaly-free model with the gauge groups $U(1)_R\times SU(3)$.
The hypermultiplets behave as a totally symmetric tensor of $SU(3)$ with 21 indices. And this model is free from the global $SU(3)$ anomaly.

\section{Summary and Discussion}

We discussed consistency conditions of six-dimensional gauged supergravity
coming from anomaly cancellation.
By performing a computer-aided search for consistent models,
we found an enormous number of anomaly-free models where
the one-loop anomaly from the fermions is
cancelled via the Green-Schwarz mechanism.

In the literature, it has often been considered
that anomaly-free models of six-dimensional gauged supergravity are quite rare.
Our results suggest that there are a huge number of other perturbatively anomaly-free models 
in six-dimensional gauged supergravity.
However, our search was limited to the cases where the gauge group is a product of $U(1)$ and/or $SU(2)$.
In fact, it is still very hard to find consistent models whenever the gauge groups consist of more than two simple Lie groups.
Thus, the existence of $E_7\times E_6 \times U(1)_R$, $E_7\times G_2 \times U(1)_R$, and $F_4 \times Sp(9) \times U(1)_R$
models found in \cite{RSSS85,AKR05,AK05} is indeed miraculous.

If one incorporates several tensor multiplets at the cost of 
covariant Lagrangian formulation,
one can employ the generalized Green-Schwarz mechanism. 
Then, if rank $\beta \le n_T + 1$, 
one can successfully cancel the local anomaly.
Thus, we might be able to find enormously many consistent models
with the gauge groups like $G_R \times G_1 \times \cdots \times G_{n_H}$ in a similar manner.
The need for the quantum formulation is much more pressing with $n_T>1$,
since in this case we cannot tell anything about the effective action in a strict sense.

We would like to comment on possible applications of our results.
We have shown several examples of anomaly-free models of $d=6,{\cal N}=(1,0)$ gauge supergravity. And some of them look very simple compared to the consistent models known so far. 
We hope that they will help to study various aspects 
of six-dimensional supergravity.

For example, when one wants to derive 
six-dimensional gauged supergravity 
from the compacitifications of type II theory 
on a smooth space, we often have only abelian gauge groups
except for $R$-symmetry,
as well as lots of drone $U(1)$'s.
If such compactification is consistent as type II string theory, 
then it should be automatically anomaly-free. 
Thus, our solutions with local $R$-symmetry $\times$ abelian factors
seem to be a good step in this direction.
However, how to obtain a large number of charged hypermultiplets from string theory and how to make gravitini charged
still remain as big problems.

Compactification to four dimensions is worth a further investigation. 
Our models might find a use in constructing higher-dimensional 
models of phenomenology and cosmology \cite{ABCFPQTZ03, BR04, ABPQ02, ABPQ03, WY02}. 
Moreover, if we compactify the theory down to four dimensions with branes \cite{ABPQ03}, 
new anomaly possibly arises on the branes.
Then one should take care of anomaly inflow \cite{CH85,GHM96}
in that framework.

Furthermore, our results may also be interesting in building solutions of $d=6,{\cal N}=(1,0)$ gauged supergravity.
For example, see the recent paper \cite{PTZ05}.

Finally, the physics of diagonally-gauged models can be studied more thoroughly. 
We may find interesting generalization of the aforementioned applications.
We hope to revisit this problem in the future.

\section*{Acknowledgements}

The authors would like to thank S. Avramis and S. Randjbar-Daemi for interesting comments and Y. Imamura, T. Eguchi for helpful discussion.
Research of one of the authors Y.T. is supported by the JSPS Fellowships for Young Scientists. 
The work is partially supported by the National Science Foundation under Grant No. PHY99-07949.

\appendix

\section{Representation Theoretical Constants}\label{appA}

Let $F_G = F^i_G T^i_r$ be the field strength of gauge group $G$, 
acting on fermions in the  representation $r$. 
When $G$ is a simple non-abelian gauge group, 
we define group-theoretical constants $C_2 (R;G),\ A (R;G)$ and $B (R;G)$ as
\begin{equation}
\tr_R F_G^2 \defeq C_2 (R;G) \, \tr_f F_G^2\,,\quad
\tr_R F_G^4 \defeq A (R;G) \, \tr_f F_G^4 + B (R;G) \pare{\tr_R F_G^2}^2\,.  \label{C2AB}
\end{equation}
where $f$ is the smallest nontrivial irreducible representation of $G$.
For exceptional-type representations which has no fourth-order Casimir invariants,
we also define $C_4 (R;G)$ as
\begin{equation}
\tr_R F_G^4 \defeq C_4 (R;G) \pare{\tr_f F_G^2}^2 = B (R;G) \, C_2 (R;G)^2 \pare{\tr_f F_G^2}^2 \,.  \label{C4}
\end{equation}
We will omit $G$ if it causes no ambiguity.

Some comments on the group-theoretical constants defined here: 
First, the ratio $\tr_r (F_G)^n / \tr_{r'} (F_G)^n$ is independent of the normalization of $T_r^i$\,.
Thus the quantities $A(R;G)$, $B(R;G)$ and $C_i(R;G)$ 
are determined only by the representation $R$ of $G$.
Second, when $R$ is the direct sum of irreducible representations $R = \oplus_i R_i$\,, 
then $C_2 (R)$ and $C_4 (R)$ are equal to 
the sums $\sum_i C_2 (R_i)$ and $\sum_i C_4 (R_i)$\,, respectively.
Third, for an irreducible exceptional-type representation $R$, we have a formula \cite{Okubo78}
\begin{equation}
C_4 (R; G) = \frac{\dim G}{2 \, \dim R \pare{2 + \dim G} } \pare{6 - \frac{C_2 ({\rm ad})}{\dim G} \cdot \frac{\dim R}{C_2 (R)}} C_2 (R; G)^2.
\end{equation}

\section{Global gauge anomaly for Majorana Weyl fermions}\label{appB}

If $\pi_6 \pare{H} = \mathbb{Z}_p$ for some gauge group $H$, 
global anomaly may exist. 
Then we must check whether the global gauge anomaly cancels. 

\subsection{Weyl fermions}

First let us review the calculation for Weyl fermions \cite{EN84,Kiritsis86,Tosa,BV97}.
The basic strategy is to embed $H$ into $G$ such that 
$\pi_7 \pare{G} = \bZ $ and $\pi_6(G)=0$. 
Then, because the gauge group $G$ has no global gauge anomaly in six dimensions, the global gauge transformation in $H$ can be deformed 
continuously to identity in $G$. 
In this way we can reduce the calculation of global anomaly for $H$ to that of perturbative anomaly for $G$.

The embedding
\begin{equation}
0 \longrightarrow H \stackrel{\iota} \longrightarrow G \stackrel{p} \longrightarrow G/H \longrightarrow 0
\end{equation}
induces the homotopy exact sequence
\begin{equation}
\cdots \stackrel{\iota_*} \longrightarrow \pi_7(G) \stackrel{p_*} \longrightarrow \pi_7(G/H) \stackrel{\partial_*} \longrightarrow \pi_6(H)\longrightarrow \pi_6(G) = 0.
\end{equation}
Let us denote by $g$, $g'$ the generators of $\pi_7(G)$ and $\pi_7(G/H)$, respectively.
Then, $ \tilde g \equiv \partial_* g' $ is a generator of $\pi_6(H)$ and there is an integer $s$ such that $p_* (g)=(g')^s$ in our cases.

Let us embed the fermion in the representation $r_L$ of $H$
in the representation $R_L \ominus R_R$ of $G$,\footnote{
Here the subscripts $L$ and $R$ denote the chirality.}
so that $R_L\ominus R_R$ decompose under $H$
to $r_L$ plus some fermions which can be massive.
Then, following the argument of \cite{EN84}, the $H$ gauge transformation corresponding to $\tilde g$ produces a phase $e^{i\theta (r)}$, with $\theta(r)$ given by
\begin{equation}
\theta(r) = \frac1s \int_{S^7} \gamma(g,A,F;R_L\ominus R_R) 
\end{equation}
where $\gamma(g,A,F;R)$ is the change under $g$ of  the 
non-abelian Chern-Simons terms in the representation $R$.
We can easily show that  $ \int_{S^7} \gamma(g,A,F;R)= 2\pi A(R;G) $
if $G=SU(n)$.
Thus we have
\begin{equation}
\theta(r)=2\pi A(R;G)/s. \label{gl_gamma}
\end{equation}

For $H = SU(2)$, $SU(3)$, and $\lie{G}_2$\,,
we can choose $G$ as $SU(4)$, $SU(4)$, $SU(7)$, respectively \cite{Kiritsis86}.
We claim that, for the representation $R$ of $SU(4)$ or $SU(7)$, $A (R; G)$ is given by
\begin{align}
A (R; SU(4)) &\equiv 2 \sum_i C_4 (r_i;H) \quad ({\rm mod}\ 6) \quad {\rm for}\ 
H = SU(2) \ {\rm or}\ SU(3)  \label{claimA}; \\
A (R; SU(7)) &\equiv 4 \sum_i C_4 (r_i;H) \quad ({\rm mod}\ 3) \quad {\rm for}\ 
H = \lie{G}_2  \label{claimG}
\end{align}
provided that 
the representation $R$ decomposes as $R = \oplus_i r_i$ under $H$\,.

To prove them, we evaluate $\tr_G F_R^4$ in two ways. Using \eqref{C2AB}, it can be rewritten as
\begin{align}
\left. \tr_G F_R^4 \right|_{{\rm on}\, H} &= A (R;G) \, 
\left. \tr_G F_f^4 \right|_{{\rm on}\, H} + B (R; G) \, 
\big(\left. \tr_G F_f^2 \right|_{{\rm on}\, H} \big)^2 \notag \\[1mm]
&= \big\{B (f;H) A (R;G) + B (R;G) \big\} \, \big(\left. \tr_G F_f^2 \right|_{{\rm on}\ H}\big)^2 . \label{evalG}
\end{align}
where $f$ is the fundamental representation of $G$, and in the last line we evaluate the trace after restricting it on $H$. 
Using the direct product decomposition $R = \oplus_i r_i$\,, the trace is
\begin{equation}
\left. \tr_G F_R^4 \right|_{{\rm on}\, H} = \tr_H \big( \sum_i F_{r_i})^4 = \sum_i \tr_H F_{r_i}^4 = \sum_i C_4 (r_i; H) \pare{\tr_H F_f^2}^2\,.
\end{equation}
By comparing the two, we get
\begin{equation}
B (f;H) A (R;G) + B (R;G) = \sum_i C_4 (r_i;H) \,,
\end{equation}
We have $B (f;H) = 1/2$ for $H = {\sf A_1, A_2}$ and $1/4$ for $H = {\sf G_2}$\,. 
Furthermore, one can show that $B (R;G) \equiv 0 \pmod{3}$ by mathematical induction \cite{OZTM87}, and the claims \eqref{claimA} and \eqref{claimG} immediately follow.
It is easy to derive the equations \eqref{g_anm1}, \eqref{g_anm3} and \eqref{g_anm5} from these results.

\subsection{Majorana Weyl fermions}
Let us now move on to the case with Majorana Weyl fermions.
As discussed in section \ref{sec2.1}, in six dimensions we can halve the degrees of freedom of Weyl spinors when they form a pseudoreal representation of the gauge groups.
Such Majorana Weyl fermions are more specifically called
symplectic Majorana Weyl fermions, though we use the two words interchangeably.

If we carry out this procedure for a hypermultiplet,
the resulting multiplet is called a half-hypermultiplet.
The gravitini, tensorini and gaugini of $d=6, {\cal N}=(1,0)$ supergravity are all Majorana Weyl, where we use the fact the $\rep{2}$ of
$Sp(1)_R$ is pseudoreal.  

Of the gauge groups which have global anomaly in $d=6$, only $SU(2)$ has
pseudoreal irreducible representations. 
So hereafter we restrict our attention to global $SU(2)$ (or $Sp(1)$) anomaly.
We assume that the perturbative anomaly is
already canceled by the Green-Schwarz mechanism.
As we saw in the preceding subsection, the Weyl fermions in $\rep{2}$
produces the phase $e^{2\pi i/6}$ under the generator of $\pi_6(SU(2))$.
Let $\alpha$ be the phase produced by Majorana Weyl fermions
in $\rep{2}$. 
Because $\alpha^2=e^{2\pi i/6}$, $\alpha$ must be either $e^{2\pi i /12}$ or $e^{2\pi i 7/12}$. Now we are going to 
determine which is the case.

To do it, we need to embed a symplectic Majorana Weyl
fermion in $\rep{2}$ of $SU(2)$ in a 
Majorana Weyl fermion in a larger gauge group without global anomaly. Thus $\rep4$ in $Sp(2)$ is a good choice. Let it decompose into
$\rep{2}\oplus \rep{1}\oplus\rep{1}$ under $SU(2)$.
We write the change of the Chern-Simons seven-form by $\gamma(g,A,F)$, as in the preceding subsection.

The phase change for Weyl fermions in the fundamental of $SU(4)$ is 
$\int\gamma(g,A,F)=2\pi$
under the generator $g$ of $\pi_7(SU(4))$.
Consider the homotopy exact sequence\footnote{
A considerable knowledge of algebraic topology is required for an actual calculation of homotopy groups.
A concise table for the higher homotopy groups of the compact Lie groups can be found in the appendix A of \cite{EDM}. 
Interested readers can consult the textbooks \cite{Hu,MT} and references therein. 
}
\begin{multline}
\pi_8(S^5)=\bZ_{24} \stackrel{\partial_*} \longrightarrow
\pi_7(Sp(2))=\bZ \stackrel{\iota_*} \longrightarrow \\
\pi_7(SU(4))=\bZ \stackrel{p_*} \longrightarrow
\pi_7(S^5)=\bZ_2 \stackrel{\partial_*} \longrightarrow
\pi_6(Sp(2))=0.
\end{multline}
It implies that the generator $g'$ of $\pi_7(Sp(2))$ is mapped to $g^2$. Thus, the phase change for Weyl fermions in $\rep{4}$ of $Sp(2)$ under $g'$ is $4\pi$.
Therefore it is $2\pi$ for Majorana Weyl fermions in $\rep{4}$.

Now consider another sequence 
\begin{multline}
\pi_7(Sp(1))=\bZ_2 \stackrel{\iota_*} \longrightarrow
\pi_7(Sp(2))=\bZ \stackrel{p_*} \longrightarrow \\
\pi_7(Sp(2)/Sp(1))=\bZ \stackrel{\partial_*} \longrightarrow
\pi_6(Sp(1))=\bZ_{12} \stackrel{\iota_*} \longrightarrow
\pi_6(Sp(2))=0.
\end{multline}
Denote the generator of $\pi_7(Sp(2)/Sp(1))$ by $h'$.
Then $h'$ satisfies $p_*(g') = (h')^{12}$, and
$\tilde h \defeq\partial_* h' $ is one of 
the generators of $\pi_6(Sp(1))$.
Thus, the phase change under $\tilde h$ for 
Majorana Weyl fermions in $\rep{2}$ is $e^{2\pi i /12}$.

Let us go on to other representations.
Let $[k]$ be the $k$-index symmetric tensor representation of $Sp(1)$ or $Sp(2)$.
Let us bear in mind that $\rep{k}=[k-1]$ in $Sp(1)$.
Then, for $Sp(2)$
\begin{equation}
\tr_{[k]}F^4=A(k)\tr_{[1]}F^4+\cdots
\end{equation}
where $A(k)=k(k+1)(k+2)(k+3)(k+4)(k^2+4k+2)/840$.
Furthermore, 
\begin{equation}
[k]\to [k]+ 2[k-1]+ 3[k-2]+4[k-3]+\cdots.
\end{equation}
under the restriction of groups from $Sp(2)$ to $Sp(1)$.
Thus,  $[k-1]_L-2[k-2]_R+[k-3]_L$ of $Sp(2)$ reduces to $\rep{k}_L$ of $Sp(1)$,
and hence the phases under the global gauge transformation $\tilde h$ for Majorana Weyl $\rep{k}$ is $2\pi D_4(\rep{k})/12$, where
\begin{equation}
D_4(\rep{k})=A(k-1)-2A(k-2)+A(k-3).
\end{equation}

Specifically, Majorana Weyl fermions in $\rep{4}$ contribute $e^{-\pi i /3}$,
and Weyl fermions in  $\rep{3}$ contribute $e^{4\pi i /3}$ to the global anomaly phase.

Finally, by considering
\begin{align*}
I_{3/2}&=\cdots+\frac {10}{3}\tr F^4, &
I_{1/2}&=\cdots+\frac 2{3}\tr F^4,
\end{align*}
and the embedding of gravitini into $\rep{4}$ of $Sp(2)$
with the symplectic Majorana Weyl condition,
we see that a gravitino contributes five times as much as that of 
a spin 1/2 fermion.

Suppose we gauge the $Sp(1)\ R$-symmetry,
then the contributions from various fermions are summarized as
\begin{equation}
\begin{array}{lrc}
\text{gravitini in a supergravity multiplet}, &  5\mod 12\,; \\
\text{tensorini in a tensor multiplet}, & -1\mod 12\,; \\
\text{the $Sp(1)_R$ gaugini},& -1 \mod 12\,; \\
\text{other gaugini in a vector multiplet}, & 1\mod 12\,.
\end{array}
\end{equation}
Thus, the condition for the cancellation of global $Sp(1)_{R(+)}$ anomaly is
\begin{equation}
n_V - D_4(\rho^H ; Sp(1))\equiv 0 \pmod{12}.
\end{equation}

%\newpage


\begin{thebibliography}{99}


    \bibitem{NS86}

  H.~Nishino and E.~Sezgin,
   ``The Complete $\mathcal{N}=2$, D = 6 Supergravity with Matter And Yang-Mills
  Couplings,''
 {\slshape   Nucl.\ Phys.\ B }{\bf 278} (1986) 353.
  %%CITATION = NUPHA,B278,353;%%
 

    \bibitem{NS97}

  H.~Nishino and E.~Sezgin,
  ``New couplings of six-dimensional supergravity,''
 {\slshape   Nucl.\ Phys.\ B }{\bf 505} (1997) 497
  [arXiv:hep-th/9703075].
  %%CITATION = HEP-TH 9703075;%%
 

    \bibitem{SS84}

  A.~Salam and E.~Sezgin,
   ``Chiral Compactification on Minkowski  $\times\ S^2$  of $\mathcal{N}=2$ Einstein-Maxwell
  Supergravity in Six-Dimensions,''
 {\slshape   Phys.\ Lett.\ B }{\bf 147} (1984) 47.
  %%CITATION = PHLTA,B147,47;%%
 

    \bibitem{GLPS03}

  R.~G\"uven, J.~T.~Liu, C.~N.~Pope and E.~Sezgin,
  ``Fine tuning and six-dimensional gauged $\mathcal{N}=(1,0)$ supergravity vacua,''
 {\slshape   Class.\ Quant.\ Grav.\  }{\bf 21} (2004) 1001
  [arXiv:hep-th/0306201].
  %%CITATION = HEP-TH 0306201;%%
 

    \bibitem{ABCFPQTZ03}

  Y.~Aghababaie {\it et al.},
  ``Warped brane worlds in six dimensional supergravity,''
 {\slshape   JHEP }{\bf 0309} (2003) 037
  [arXiv:hep-th/0308064].
  %%CITATION = HEP-TH 0308064;%%
 

    \bibitem{BR04}

  R.~H.~Brandenberger and S.~Randjbar-Daemi,
  ``Inflation in 6D gauged supergravity,''
 {\slshape   JHEP }{\bf 0410} (2004) 022
  [arXiv:hep-th/0404228].
  %%CITATION = HEP-TH 0404228;%%
 

    \bibitem{AW83}

  L.~Alvarez-Gaum\'e and E.~Witten,
  ``Gravitational Anomalies,''
 {\slshape   Nucl.\ Phys.\ B }{\bf 234} (1984) 269.
  %%CITATION = NUPHA,B234,269;%%
 

    \bibitem{Ketov88}

  S.~V.~Ketov,
  ``Anomalies in 6-dimensional gauge theories,''
 {\slshape   Sov.\ J.\ Nucl.\ Phys.\  }{\bf 47} (1988) 943
  [Yad.\ Fiz.\  {\bf 47} (1988) 1484].
  %%CITATION = SJNCA,47,943;%%
 

    \bibitem{Ketov90}

  S.~V.~Ketov,
  ``Anomalies of Kaluza-Klein Theories in Six-Dimensions,''
 {\slshape   Class.\ Quant.\ Grav.\  }{\bf 7} (1990) 1387.
  %%CITATION = CQGRD,7,1387;%%
 

    \bibitem{GSW85}

  M.~B.~Green, J.~H.~Schwarz and P.~C.~West,
  ``Anomaly Free Chiral Theories in Six-Dimensions,''
 {\slshape   Nucl.\ Phys.\ B }{\bf 254} (1985) 327.
  %%CITATION = NUPHA,B254,327;%%
 

    \bibitem{Schwarz95}

  J.~H.~Schwarz,
  ``Anomaly-Free Supersymmetric Models in Six Dimensions,''
 {\slshape   Phys.\ Lett.\ B }{\bf 371} (1996) 223
  [arXiv:hep-th/9512053].
  %%CITATION = HEP-TH 9512053;%%
 

    \bibitem{SW96}

  N.~Seiberg and E.~Witten,
  ``Comments on String Dynamics in Six Dimensions,''
 {\slshape   Nucl.\ Phys.\ B }{\bf 471} (1996) 121
  [arXiv:hep-th/9603003].
  %%CITATION = HEP-TH 9603003;%%
 

    \bibitem{GS84}

  M.~B.~Green and J.~H.~Schwarz,
   ``Anomaly Cancellation in Supersymmetric D=10 Gauge Theory And Superstring
  Theory,''
 {\slshape   Phys.\ Lett.\ B }{\bf 149} (1984) 117.
  %%CITATION = PHLTA,B149,117;%%
 

    \bibitem{RSSS85}

  S.~Randjbar-Daemi, A.~Salam, E.~Sezgin and J.~A.~Strathdee,
  ``An Anomaly Free Model in Six-Dimensions,''
 {\slshape   Phys.\ Lett.\ B }{\bf 151} (1985) 351.
  %%CITATION = PHLTA,B151,351;%%
 

    \bibitem{AKR05}

  S.~D.~Avramis, A.~Kehagias and S.~Randjbar-Daemi,
  ``A new anomaly-free gauged supergravity in six dimensions,''
 {\slshape   JHEP }{\bf 0505} (2005) 057
  [arXiv:hep-th/0504033].
  %%CITATION = HEP-TH 0504033;%%
 

    \bibitem{AK05}

  S.~D.~Avramis and A.~Kehagias,
  ``A systematic search for anomaly-free supergravities in six dimensions,''
 {\slshape   JHEP }{\bf 0510} (2005) 052
  [arXiv:hep-th/0508172].
  %%CITATION = HEP-TH 0508172;%%
 

    \bibitem{CGP03}

  M.~Cveti\v c, G.~W.~Gibbons and C.~N.~Pope,
  ``A string and M-theory origin for the Salam-Sezgin model,''
 {\slshape   Nucl.\ Phys.\ B }{\bf 677} (2004) 164
  [arXiv:hep-th/0308026].
  %%CITATION = HEP-TH 0308026;%%
 

    \bibitem{BJS05}

  E.~Bergshoeff, D.~C.~Jong and E.~Sezgin,
   ``Noncompact gaugings, chiral reduction and dual sigma models in
  supergravity,''
 {\slshape   Class.\ Quant.\ Grav.\  }{\bf 23} (2006) 2803
  [arXiv:hep-th/0509203].
  %%CITATION = HEP-TH 0509203;%%
 

    \bibitem{RS04}

  S.~Randjbar-Daemi and E.~Sezgin,
  ``Scalar potential and dyonic strings in 6d gauged supergravity,''
 {\slshape   Nucl.\ Phys.\ B }{\bf 692} (2004) 346
  [arXiv:hep-th/0402217].
  %%CITATION = HEP-TH 0402217;%%
 

    \bibitem{RSS82}

  S.~Randjbar-Daemi, A.~Salam and J.~A.~Strathdee,
  ``Spontaneous Compactification in Six-Dimensional Einstein-Maxwell Theory,''
 {\slshape   Nucl.\ Phys.\ B }{\bf 214} (1983) 491.
  %%CITATION = NUPHA,B214,491;%%
 

    \bibitem{RSS83}

  S.~Randjbar-Daemi, A.~Salam and J.~A.~Strathdee,
  ``Instability of Higher Dimensional Yang-Mills Systems,''
 {\slshape   Phys.\ Lett.\ B }{\bf 124} (1983) 345
  [Erratum-ibid.\ B {\bf 144} (1984) 455].
  %%CITATION = PHLTA,B124,345;%%
 

    \bibitem{FMS96}

  S.~Ferrara, R.~Minasian and A.~Sagnotti,
  ``Low-Energy Analysis of $M$ and $F$ Theories on Calabi-Yau Threefolds,''
 {\slshape   Nucl.\ Phys.\ B }{\bf 474} (1996) 323
  [arXiv:hep-th/9604097].
  %%CITATION = HEP-TH 9604097;%%
 

    \bibitem{AG84}

  L.~Alvarez-Gaum\'e and P.~H.~Ginsparg,
  ``The Structure of Gauge And Gravitational Anomalies,''
 {\slshape   Annals Phys.\  }{\bf 161} (1985) 423
  [Erratum-ibid.\  {\bf 171} (1986) 233].
  %%CITATION = APNYA,161,423;%%
 

    \bibitem{Okubo78}

  S.~Okubo,
  ``Quartic Trace Identity for Exceptional Lie Algebras,''
 {\slshape   J.\ Math.\ Phys.\  }{\bf 20} (1979) 586.
  %%CITATION = JMAPA,20,586;%%
 

    \bibitem{Okubo81}

  S.~Okubo,
   ``Modified Fourth Order Casimir Invariants And Indices for Simple Lie
  Algebras,''
 {\slshape   J.\ Math.\ Phys.\  }{\bf 23} (1982) 8.
  %%CITATION = JMAPA,23,8;%%
 

    \bibitem{Witten82}

  E.~Witten,
  ``An $SU(2)$ anomaly,''
 {\slshape   Phys.\ Lett.\ B }{\bf 117} (1982) 324.
  %%CITATION = PHLTA,B117,324;%%
 

    \bibitem{EN84}

  S.~Elitzur and V.~P.~Nair,
  ``Nonperturbative Anomalies in Higher Dimensions,''
 {\slshape   Nucl.\ Phys.\ B }{\bf 243} (1984) 205.
  %%CITATION = NUPHA,B243,205;%%
 

    \bibitem{Kiritsis86}

  E.~Kiritsis,
  ``GLOBAL GAUGE ANOMALIES IN HIGHER DIMENSIONS,''
 {\slshape   Phys.\ Lett.\ B }{\bf 178} (1986) 53
  [Erratum-ibid.\ B {\bf 181} (1986) 416].
  %%CITATION = PHLTA,B178,53;%%
 

    \bibitem{ZOT87}

  H.-Z.~Zhang, S.~Okubo and Y.~Tosa,
  ``Global Gauge Anomalies for Simple Lie Algebras,''
 {\slshape   Phys.\ Rev.\ D }{\bf 37} (1988) 2946.
  %%CITATION = PHRVA,D37,2946;%%
 

    \bibitem{Tosa}

  Y.~Tosa,
   ``Global Gauge Anomalies for Theories with The Green-Schwarz Local Anomaly
  Cancellation Mechanism,''
 {\slshape   Phys.\ Rev.\ D }{\bf 40} (1989) 1934.
  %%CITATION = PHRVA,D40,1934;%%
 

    \bibitem{BV97}

  M.~Bershadsky and C.~Vafa,
   ``Global anomalies and geometric engineering of critical theories in six
  dimensions,''
  arXiv:hep-th/9703167.
  %%CITATION = HEP-TH 9703167;%%
 

    \bibitem{BKSS86}

  E.~Bergshoeff, T.~W.~Kephart, A.~Salam and E.~Sezgin,
  ``Global Anomalies in Six-Dimensions,''
 {\slshape   Mod.\ Phys.\ Lett.\ A }{\bf 1} (1986) 267.
  %%CITATION = MPLAE,A1,267;%%
 

    \bibitem{ABPQ02}

  Y.~Aghababaie, C.~P.~Burgess, S.~L.~Parameswaran and F.~Quevedo,
   ``SUSY breaking and moduli stabilization from fluxes in gauged 6D
  supergravity,''
 {\slshape   JHEP }{\bf 0303} (2003) 032
  [arXiv:hep-th/0212091].
  %%CITATION = HEP-TH 0212091;%%
 

    \bibitem{ABPQ03}

  Y.~Aghababaie, C.~P.~Burgess, S.~L.~Parameswaran and F.~Quevedo,
   ``Towards a naturally small cosmological constant from branes in 6D
  supergravity,''
 {\slshape   Nucl.\ Phys.\ B }{\bf 680} (2004) 389
  [arXiv:hep-th/0304256].
  %%CITATION = HEP-TH 0304256;%%
 

    \bibitem{WY02}

  T.~Watari and T.~Yanagida,
   ``Geometric origin of large lepton mixing in a higher dimensional
  spacetime,''
 {\slshape   Phys.\ Lett.\ B }{\bf 544} (2002) 167
  [arXiv:hep-ph/0205090].
  %%CITATION = HEP-PH 0205090;%%
 

    \bibitem{CH85}

  C.~G.~Callan and J.~A.~Harvey,
  ``Anomalies And Fermion Zero Modes on Strings And Domain Walls,''
 {\slshape   Nucl.\ Phys.\ B }{\bf 250} (1985) 427.
  %%CITATION = NUPHA,B250,427;%%
 

    \bibitem{GHM96}

  M.~B.~Green, J.~A.~Harvey and G.~W.~Moore,
  ``I-brane inflow and anomalous couplings on D-branes,''
 {\slshape   Class.\ Quant.\ Grav.\  }{\bf 14} (1997) 47
  [arXiv:hep-th/9605033].
  %%CITATION = HEP-TH 9605033;%%
 

    \bibitem{PTZ05}

  S.~L.~Parameswaran, G.~Tasinato and I.~Zavala,
  ``The 6D SuperSwirl,''
 {\slshape   Nucl.\ Phys.\ B }{\bf 737} (2006) 49
  [arXiv:hep-th/0509061].
  %%CITATION = HEP-TH 0509061;%%
 

    \bibitem{OZTM87}

  S.~Okubo, H.~Zhang, Y.~Tosa and R.~E.~Marshak,
  ``$SU(N)$ GLOBAL GAUGE ANOMALIES IN EVEN DIMENSIONS,''
 {\slshape   Phys.\ Rev.\ D }{\bf 37} (1988) 1655.
  %%CITATION = PHRVA,D37,1655;%%
 

    \bibitem{EDM}


 Mathematical Society of Japan, ed., ``Encyclopedic Dictionary of Mathematics'',  MIT press, 1987. 

    \bibitem{Hu}


 S.-T. Hu, ``Homotopy Theory'', Academic Press, 1959

    \bibitem{MT}


 M. Mimura and H. Toda, ``Topology of Lie Groups, I, II'', American Mathematical Society, 1991.


\end{thebibliography}
\end{document}